\documentclass[aps,prd,preprint,groupedaddress]{revtex4-1}
\usepackage{amsmath}
\usepackage{graphicx}

\def\CU{C_U}
\def\CD{C_D}

\def\CV{C_V}
\def\CW{C_W}
\def\CWZ{C_{WZ}}
\def\CZ{C_Z}
\def\CG{C_g}
\def\CP{C_\gamma}
\def\cu{\CU}
\def\cd{\CD}
\def\cv{\CV}
\def\cw{\CW}
\def\cz{\CZ}
\def\cg{\CG}
\def\cp{\CP}
\def\cwz{\CWZ}

\def\dcg{\Delta \CG}
\def\dcp{\Delta \CP}

\def\sina{{\rm s}_\alpha}
\def\cosa{{\rm c}_\alpha}
\def\sinb{{\rm s}_\beta}
\def\cosb{{\rm c}_\beta}
\def\tanb{\tan\beta}
\def\cotb{\cot\beta}
\def\cosba{\cos(\beta-\alpha)}
\def\sinba{\sin(\beta-\alpha)}

\def\br{{\mathcal B}}
\def\brinv{\br_{\rm inv}}
\def\brnew{\br_{\rm new}}

\begin{document}


\title{Status of Higgs couplings after Run-1 of the LHC using Lilith~1.0}

\author{J\'er\'emy Bernon}
\email[]{jeremy.bernon@lpsc.in2p3.fr}
\author{B\'eranger Dumont}
\email[]{beranger.dumont@lpsc.in2p3.fr}
\author{Sabine Kraml}
\email[]{sabine.kraml@lpsc.in2p3.fr}
\affiliation{Laboratoire de Physique Subatomique et de Cosmologie (LPSC), 
Universit\'e Grenoble-Alpes, CNRS/IN2P3, 53 Avenue des Martyrs, F-38026 Grenoble, France} 

\date{\today}

\begin{abstract}
We provide an update of the global fits of the couplings of the $125.5$~GeV Higgs boson using all publicly available experimental results from Run-1 of the LHC as per Summer 2014. The fits are done by means of the new public code {\tt Lilith~1.0}. We present a selection of results given in terms of signal strengths, reduced couplings, and for the Two-Higgs-Doublet Models of Type~I and~II.
\end{abstract}


\maketitle

\section{Introduction \label{intro}}

The properties of the observed Higgs boson with mass around 125~GeV~\cite{Aad:2012tfa,Chatrchyan:2012ufa} have been measured with unforeseeable precision already during Run-1 of the LHC 
at 7--8~TeV center-of-mass energy~\cite{ATLAS-CONF-2014-009,CMS-PAS-HIG-14-009}.
This is a consequence of the excellent operation of the LHC  and of the wealth of accessible final states for a 125~GeV Standard Model (SM)-like Higgs boson.
Indeed, many distinct signal strengths, defined as production$\times$decay rates relative to SM expectations, $\mu_i\equiv (\sigma\times\br)_i/(\sigma\times\br)_i^{\rm SM}$, have been measured and used to obtain information about the couplings of 
the Higgs boson to electroweak gauge bosons, fermions of the third generation, and 
loop-induced couplings to photons and gluons.
(See \cite{Boudjema:2013qla} for a thorough discussion of the use of signal strengths $\mu_i$.)

Fits to various combinations of reduced Higgs couplings, {\it i.e.}\ Higgs couplings to fermions and gauge bosons relative to their SM values, have been performed by the experimental collaborations themselves, {\it e.g.}, in~\cite{ATLAS-CONF-2014-009,CMS-PAS-HIG-14-009}. Moreover, theorists combine the results from ATLAS and CMS 
in global fits, see {\it e.g.}~\cite{Belanger:2013xza,Dumont:2013npa} and references therein, in order to test consistency with SM expectations and to constrain models with modified Higgs couplings. In particular, the couplings of the observed Higgs boson could deviate from the SM predictions due to the presence of other Higgs states mixing with the observed one and/or due to new particles contributing to the loop-induced couplings.  

In~\cite{Belanger:2013xza}, a comprehensive analysis of the Higgs signal strengths and couplings and implications for extended Higgs sectors was performed based on the experimental results as per Spring 2013. Since then, a number of new measurements or updates of existing ones were published by the experimental collaborations. 
From ATLAS, the $VH,\,H \to b \bar b$ and the $H\to\tau\tau$ results were updated with full luminosity~\cite{ATLAS-CONF-2013-079,ATLAS-CONF-2013-108}. Moreover, significantly improved measurements in the $H \to \gamma\gamma$~\cite{Aad:2014eha} and $H \to ZZ^*$~\cite{Aad:2014eva} channels were released, and the search for invisible decays in the $ZH \to \ell\ell + {\rm invisible}$ channel was updated~\cite{Aad:2014iia}.
There were also significant news from CMS, in particular 
updates of the $H \to ZZ^* \to 4\ell$~\cite{Chatrchyan:2013mxa} 
and $H \to WW^*$~\cite{Chatrchyan:2013iaa} results, 
and---most importantly---the long-awaited final results for the $H \to \gamma\gamma$ channels~\cite{Khachatryan:2014ira}. 
Furthermore, CMS published new results for $H\to \tau\tau$~\cite{Chatrchyan:2014nva} and 
$H\to {\rm invisible}$~\cite{Chatrchyan:2014tja}.
Finally, in both ATLAS and CMS a special effort was made for probing the production of a Higgs boson in association with a pair of top quarks (ttH). From ATLAS, ttH results are available for $H \to \gamma\gamma$~\cite{Aad:2014eha} and $H \to b\bar b$~\cite{ATLAS-CONF-2014-011}, while CMS published ttH results for $H \to \gamma\gamma, b\bar b, \tau\tau$, $ZZ^*$ and $WW^*$~\cite{Khachatryan:2014qaa}.

We therefore think that an update of the global coupling fits, combining ATLAS and CMS results, is timely and interesting for the high-energy-physics community in general, even more so as this will likely define the status of the Higgs couplings until the first  round of Higgs results will become available from LHC Run-2 (or until an official combination of the Run-1 results is done by ATLAS and CMS).  
Hence, in this short communication, we provide such an update for 
{\it i)}~the combined signal strengths, 
{\it ii)}~the most important reduced coupling fits, and 
{\it iii)}~Two Higgs Doublet Models of Type~I and Type~II 
by means of a new public code, {\tt Lilith~1.0}~\cite{lilith-webpage}.  

{\tt Lilith} stands for ``LIght LIkelihood fiT for the Higgs''. It is a light and easy-to-use {\tt Python} tool to determine the likelihood of a generic Higgs boson with mass around 125~GeV from the latest experimental data, and can conveniently be used to fit the Higgs couplings and/or put constraints on theories beyond the SM. The experimental results used are the signal strengths in the primary Higgs production modes~\footnote{The five theoretically ``pure'' production modes which are accessible are gluon--gluon fusion (ggF),  vector boson fusion (VBF), associated production with a $W$ or $Z$ boson (WH and ZH, commonly denoted as VH), and associated production with a top-quark pair (ttH).} 
as published by the ATLAS and CMS experiments at the LHC and by the Tevatron experiments. All experimental data are stored in a flexible XML database which is easy to maintain. {\tt Lilith~1.0} has been validated extensively against the ATLAS and CMS coupling fits, see~\cite{lilith-webpage}. A quick user guide is also available from~\cite{lilith-webpage}; 
a manual providing a complete description of the code is in preparation.

\section{Combined Signal Strengths \label{comb-mu}}

\begin{figure}[t!]\centering
\includegraphics[width=6cm]{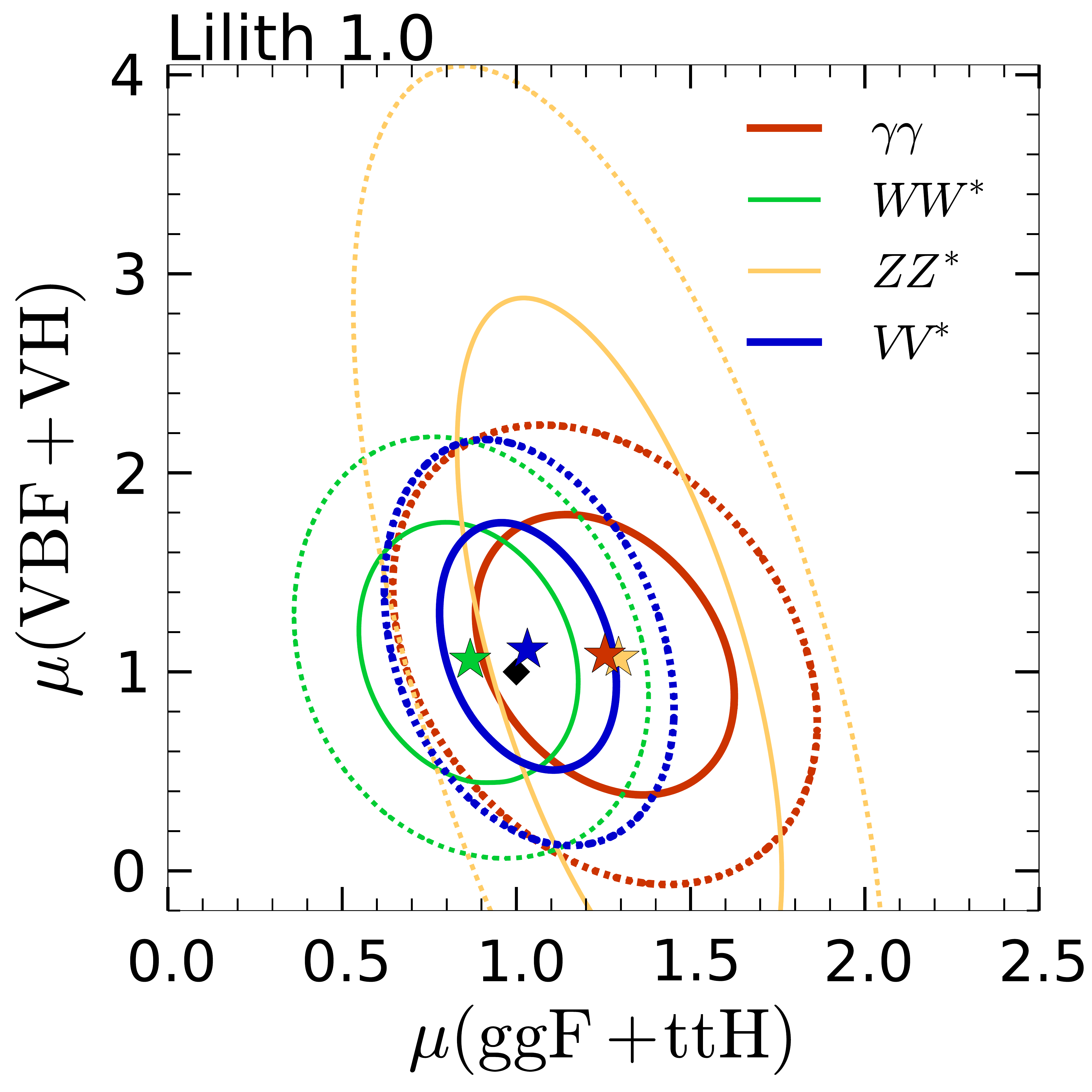}\quad 
\includegraphics[width=6cm]{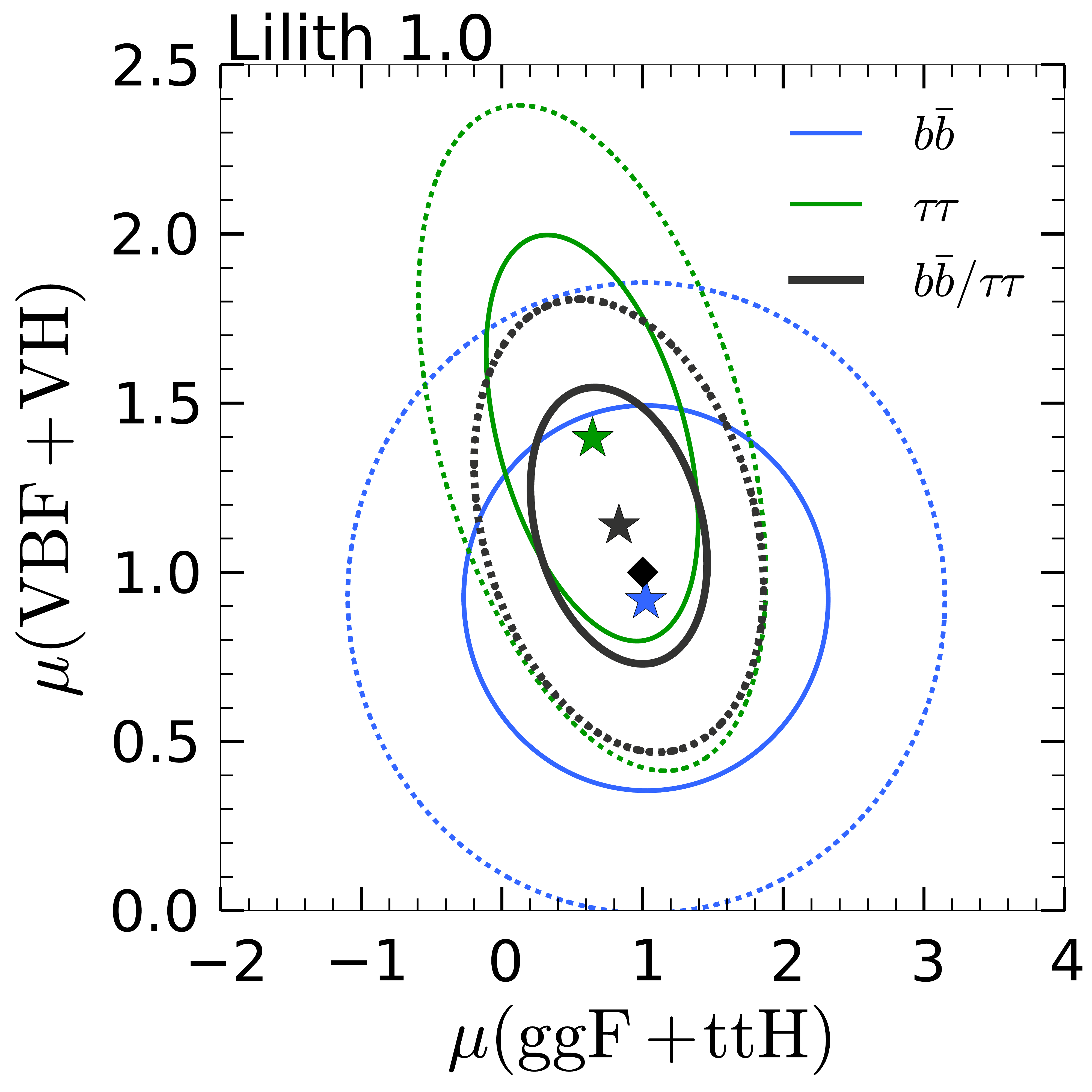}
\caption{Combined signal strengths in the plane of $({\rm ggF+ttH})$ versus $({\rm VBF+VH})$ production, 
on the left for the $\gamma\gamma$, $ZZ^*$, $WW^*$ and $VV^*$ decay modes (the latter assuming $ZZ^*=WW^*$), 
on the right for the $b\bar b$ and $\tau\tau$ decay modes and their combination $b\bar b=\tau\tau$. 
The full (dashed) contours denote the 68.3\% (95.4\%)~CL regions, derived by combining the ATLAS, CMS and Tevatron results.  The best-fit points are marked as stars, and the SM case by a black diamond. 
\label{fig:ellipses1} }
\end{figure}

We begin by showing in Fig.~\ref{fig:ellipses1} contours of constant confidence level (CL) for the 
combined signal strengths in the $\mu({\rm ggF+ttH})$ versus $\mu({\rm VBF+VH})$ plane for different 
Higgs decay modes. The left panel shows the bosonic channels $H\to\gamma\gamma,\,WW^*,\,ZZ^*$ 
as well as $VV^*$, where $VV^* \equiv ZZ^*,WW^*$;  the right panel shows the fermonic channels 
$b\bar b,\,\tau\tau$ as well as $b\bar b=\tau\tau$.

The combination of the $ZZ^*$ and $WW^*$ decay modes is justified by custodial symmetry, 
which implies that the $HZZ$ and $HWW$ couplings are rescaled by the same factor with respect to the SM. 
The combination of the $b\bar{b}$ and $\tau\tau$ decay modes is justified, in principle,
in models where one specific Higgs doublet has the same couplings, with respect
to the SM, to down-type quarks and leptons, although QCD corrections
can lead to deviations of the reduced $Hbb$ and $H\tau\tau$ couplings from a common value.

All results show an excellent agreement with the SM. 
Compared to \cite{Belanger:2013xza},  uncertainties have been significantly reduced for the fermionic channels, 
particularly for $H \to b\bar b$ in ttH production. As for $H \to \gamma\gamma$, while previously small excesses 
were observed in ggF by ATLAS and in ${\rm VBF+VH}$ by both ATLAS and (to a lesser extent) CMS, updated 
results point to a more SM-like behavior.  At the same time, the slight deficit previously seen by CMS in ggF 
is no longer present. Overall, this leads to a central value only slightly larger than unity. 

A comment is in order here.  
In the latest experimental papers, only the 68\% and 95\%~CL contours are displayed in the 
$\mu({\rm ggF + ttH})$ versus $\mu({\rm VBF + VH})$ plots, or in other 2D projections.   
In order to use this information, one is forced to make assumptions on the likelihood functions---typically this 
means assuming normally distributed signal strengths, and this is also the approach we have adopted here. 
However, this is not fully satisfactory and sometimes reproduces the contours rather poorly, 
as in the case of ATLAS $H \rightarrow ZZ^*$. (See \cite{Boudjema:2013qla} for a detailed discussion.) 
In the previous round of Higgs results, 
CMS had provided a temperature plot for the $H \to \gamma\gamma$ result~\cite{Khachatryan:2014ira}, while 
ATLAS had gone a step further and digitally published the 2D likelihood grids for the bosonic channels~\cite{ATLAS-data-Hgamgam,ATLAS-data-HZZ,ATLAS-data-HWW} corresponding to the results of~\cite{Aad:2013wqa}. 
This was a boon for interpretation studies,  as it rendered the Gaussian approximation unnecessary at least for these channels. We strongly hope that such likelihood grids (or digitized temperature plots) will again be made available in the future by both ATLAS and CMS.

\begin{table}[tb!]  
\center
\renewcommand{\arraystretch}{1.1}
\begin{tabular}{|c|c|c|c||c|c|c|}
\hline
& $\widehat{\mu}^{\rm{ggF}}$ & $\widehat{\mu}^{\rm{VBF}}$ & $\rho$ & $a$ & $b$ & $c$ \\
\hline 
$\gamma\gamma$ & $1.25 \pm 0.24$ & $1.09 \pm 0.46$ & ~$-0.30$~ & ~$18.26$~ & ~2.84~ & ~\phantom{0}5.08~ \\
\hline
$VV^*$ & $1.03 \pm 0.17$ & $1.12 \pm 0.41$ & $-0.29$ & ~39.07~ & 4.68 & \phantom{0}6.52 \\
\hline
$ZZ^*$ & $1.30\pm 0.31$ & $1.06\pm 1.20$ & $-0.59$ & 16.27 & 2.45 & \phantom{0}1.06 \\
\hline
$WW^*$ & $0.86\pm 0.21$ & $1.09\pm 0.43$ & $-0.20$ & 24.15 & 2.29 & \phantom{0}5.58 \\
\hline \hline
~$b\bar{b}/\tau\tau$~ & $0.83 \pm 0.41$ & $1.14 \pm 0.27$ & $-0.27$ & \phantom{0}6.29 & 2.62 &
~14.86~ \\
\hline
$b\bar{b}$ & ~$1.02 \pm 0.85$~ & ~$0.92 \pm 0.38$~ & 0 & \phantom{0}1.37 & 0 & \phantom{0}7.10 \\
\hline
$\tau\tau$ & $0.64 \pm 0.50$ & ~$1.40 \pm 0.40$~ & $-0.42$ & \phantom{0}4.92 & 2.60 & \phantom{0}7.76 \\
\hline
\end{tabular}
\caption{Combined best-fit signal strengths $\widehat{\mu}^{\rm{ggF}}$, $\widehat{\mu}^{\rm{VBF}}$ 
and correlation coefficient $\rho$ for various Higgs decay modes (with $VV^* \equiv WW^*,\,ZZ^*$), as well as the coefficients 
$a$, $b$ and $c$ for the approximate $\chi^2$ in Eq.~(\ref{eq:1}).}
\label{tab:1}
\end{table}

In the Gaussian approximation, we can derive a simple expression for the $\chi^2$ for each decay mode $j$ 
in the form of ellipses \cite{Belanger:2013xza}
%
%
\begin{equation}
\chi_j^2  = a_j(\mu_j^{\rm{ggF}}-\widehat{\mu}_j^{\rm{ggF}})^2 
   + 2b_j(\mu_j^{\rm{ggF}}-\widehat{\mu}_j^{\rm{ggF}})(\mu_j^{\rm{VBF}}-\widehat{\mu}_j^{\rm{VBF}})  
   + c_j(\mu_j^{\rm{VBF}}-\widehat{\mu}_j^{\rm{VBF}})^2 \,\\
\label{eq:1}
\end{equation}
%
%
%
where the upper indices ggF and VBF stand for $({\rm ggF+ttH})$ and $({\rm VBF+VH})$, respectively, 
and $\widehat{\mu}_j^{\rm{ggF}}$ and $\widehat{\mu}_j^{\rm{VBF}}$ denote the
best-fit points obtained from the measurements. 
The parameters $\widehat{\mu}^{\rm{ggF}}$, $\widehat{\mu}^{\rm{VBF}}$, $a$, $b$ and
$c$ for Eq.~(\ref{eq:1}) (and, for completeness, the correlation coefficient $\rho$) 
resulting from our fit are listed in Table~\ref{tab:1}. 
Approximating the $\chi^2$ in this form can be useful for applications that aim at a 
quick assessment of the compatibility with the experimental data without invoking the complete likelihood calculation. 
In the fits presented below, we will apply the full machinery of {\tt Lilith~1.0}.

\clearpage 
\section{Fits to reduced Higgs couplings}

Let us now turn to the fits of reduced couplings. 
To this end, we define 
%
%
\begin{equation}
\mathcal{L}=\left[C_W m_W W^\mu W_\mu + C_Z \frac{m_Z}{\cos \theta_W} Z^\mu Z_\mu 
  - C_U \frac{m_t}{2m_W} \bar{t}t - C_D \frac{m_b}{2 m_W} \bar{b}b - C_D \frac{m_\tau}{2 m_W} \bar{\tau}\tau \right]H \,,  
\label{eq:hfit-HL}
\end{equation}
%
%
%
where the $C_I$ are scaling factors for the couplings relative to their SM values, introduced to test possible deviations in the data from SM expectations.  
We set $C_W, C_Z > 0$ by convention; 
custodial symmetry implies $\CV\equiv\CW=\CZ$. 

In addition to these tree-level couplings, we define the loop-induced 
couplings $C_g$ and $C_\gamma$ of the $H$ to $gg$ and $\gamma\gamma$, respectively. 
With the {\tt BEST-QCD} option in {\tt Lilith~1.0}, the contributions of SM particles to 
$C_g$ and $C_\gamma$ (as well as the corrections to VBF production) are computed at NLO QCD 
from the given values for $\cu$, $\cd$, $\cw$ and $\cz$ 
following the procedure recommended by the  
LHC Higgs Cross Section Working Group~\cite{LHCHiggsCrossSectionWorkingGroup:2012nn} 
(using grids generated from  \texttt{HIGLU}~\cite{Spira:1995mt}, 
\texttt{HDECAY}~\cite{Djouadi:1997yw}, and {\tt VBFNLO}~\cite{Arnold:2011wj}). 
Alternatively, $C_g$ and $C_\gamma$ can be taken as free parameters. 
Finally, invisible or undetected branching ratios can also be included in the fit.

Deviations from SM expectations can be divided into two categories: 1.~modifications of the tree-level couplings, as in extended Higgs sectors or Higgs portal models, and 2.~vertex loop effects from new particles beyond the SM, modifying in particular $\cg$ and/or $\cp$. We first discuss the former.

\begin{figure}[t!]\centering
\includegraphics[width=6.5cm]{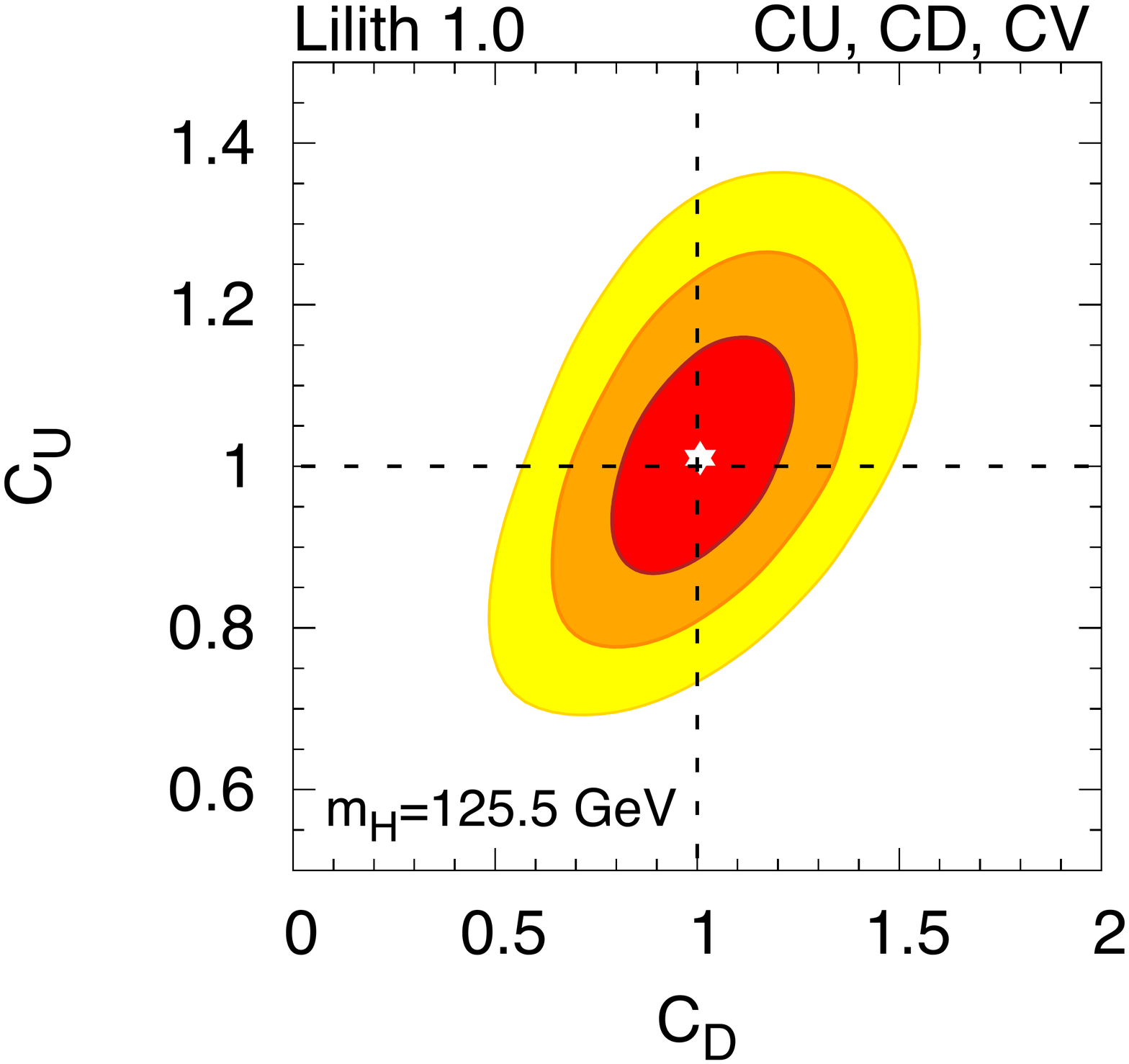}\hspace*{-1.2cm} 
\includegraphics[width=6.5cm]{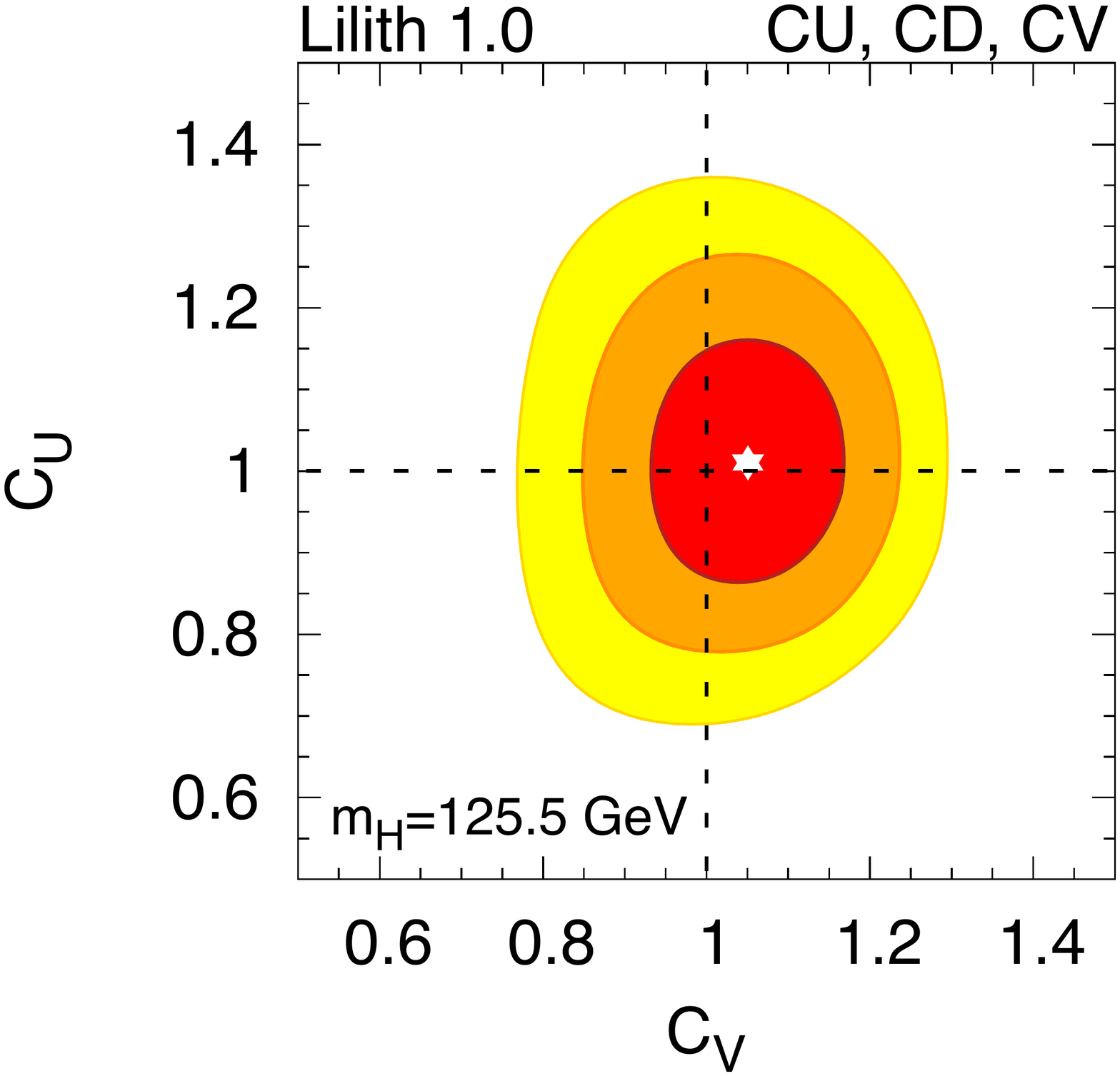}\hspace*{-1cm}
\includegraphics[width=6.5cm]{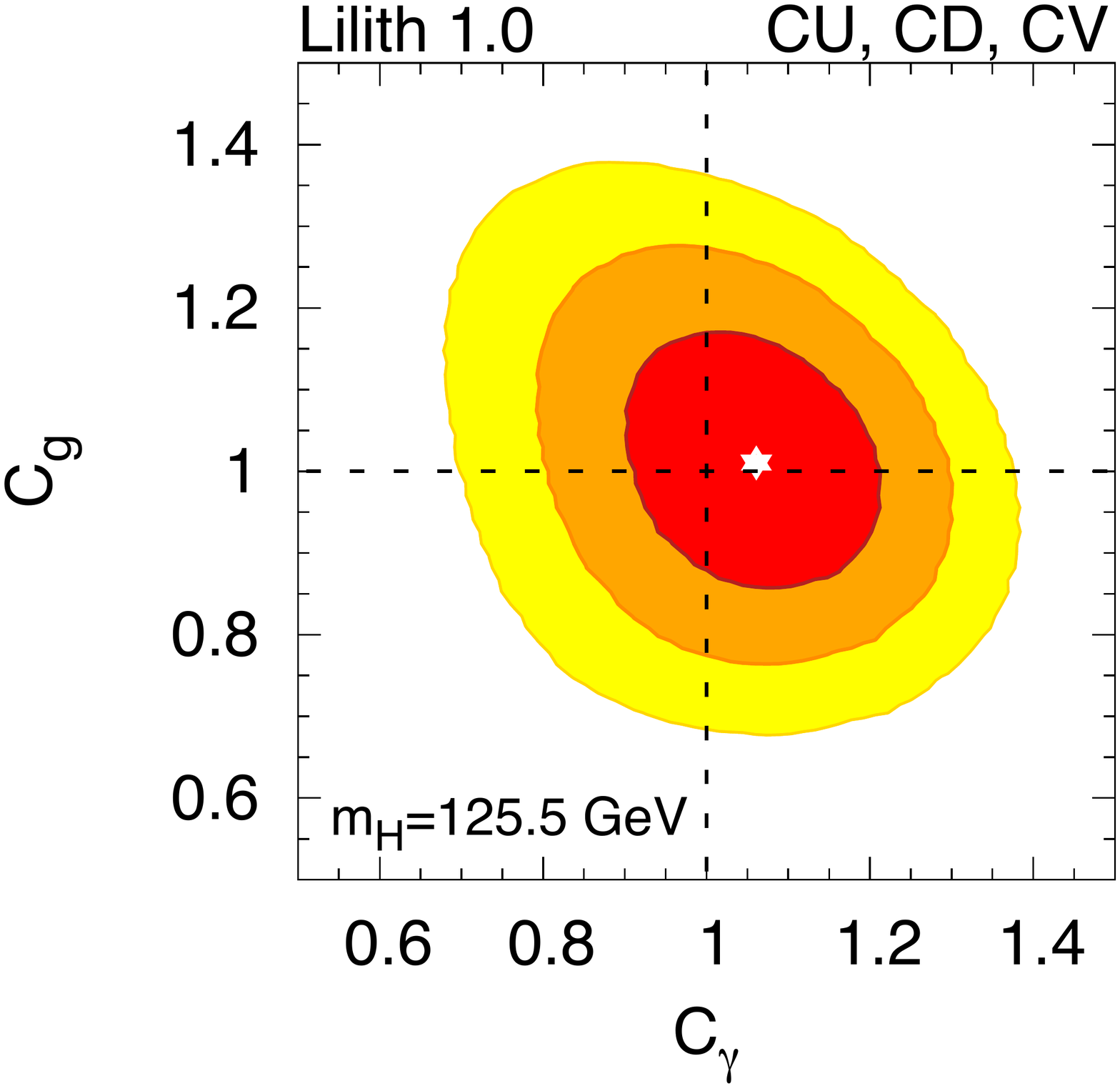} 
\caption{Fits of $\CU$, $\CD$ and $\CV$ (left and middle panels) and resulting $C_g$ versus $C_\gamma$ (right panel). 
The red, orange and yellow areas are the 68.3\%, 95.4\% and 99.7\%~CL regions, respectively.  
The best-fit points are marked as white stars. 
Invisible or undetected decays are assumed to be absent. 
\label{fig:cu-cd-cv} }
\end{figure}

Figure~\ref{fig:cu-cd-cv} shows results for a 3-parameter fit of $\CU$, $\CD$, $\CV$, 
assuming custodial symmetry and taking $\CU,\,\CD>0$. 
We note that at $95.4\%$ CL in 2D, $\cu$ and $\cv$ are constrained within roughly $\pm 20\%$; 
the uncertainty on $\cd$ is about twice as large.   
Although not shown in Fig.~\ref{fig:cu-cd-cv}, $\cu<0$ is excluded at more than $2\sigma$, 
while the sign ambiguity in $\cd$ remains. 
(See \cite{Ferreira:2014naa} for a discussion of wrong-sign Yukawa couplings.) 
The fact that $\widehat\mu^{\rm ggF}_{\gamma\gamma}$, $\widehat\mu^{\rm VBF}_{\gamma\gamma}$ 
and $\widehat\mu^{\rm VBF}_{VV}$ lie somewhat above one (cf.\ Fig.~\ref{fig:ellipses1} and Table~\ref{tab:1})  
leads to a slight preference for $C_V>1$. 
The best fit is obtained for $\CU=\CD=1.01$ and $\CV=1.05$, resulting in $\CG=1.01$ and $\CP=1.06$. 
All these reduced couplings are however consistent with unity at the 1$\sigma$ level.
In 1D, {\it i.e.}\ profiling over the other parameters, we find 
$\cu=[0.91,1.11]$ ($[0.82,1.22]$), $\cd=[0.85,1.16]$ ($[0.70,1.32]$), and  
$\cv=[0.97,1.13]$ ($[0.89,1.20]$) at 68.3\% (95.4\%)~CL; 
requiring $\cv<1$, we get $\CV> 0.96$ (0.88). 

To test possible deviations from custodial symmetry, we next define $\cwz\equiv \cw/\cz$ and perform a 4-parameter fit of $\cu,\cd, \cz, \cwz$. 
In 1D, we find $\cwz=[0.83,1.02]$ ($[0.75, 1.16]$) and $\cz=[1.0,1.24]$ ($[0.89,1.35]$) at 68.3\% (95.4\%)~CL.  
(The corresponding 68.3\% and 95.4\%~CL intervals for $\cw$ are $[0.95, 1.11]$ and $[0.87, 1.19]$.)
Current Higgs data hence provide a significant constraint on deviations from custodial symmetry. 

\begin{figure}[t!]\centering
\includegraphics[width=6.5cm]{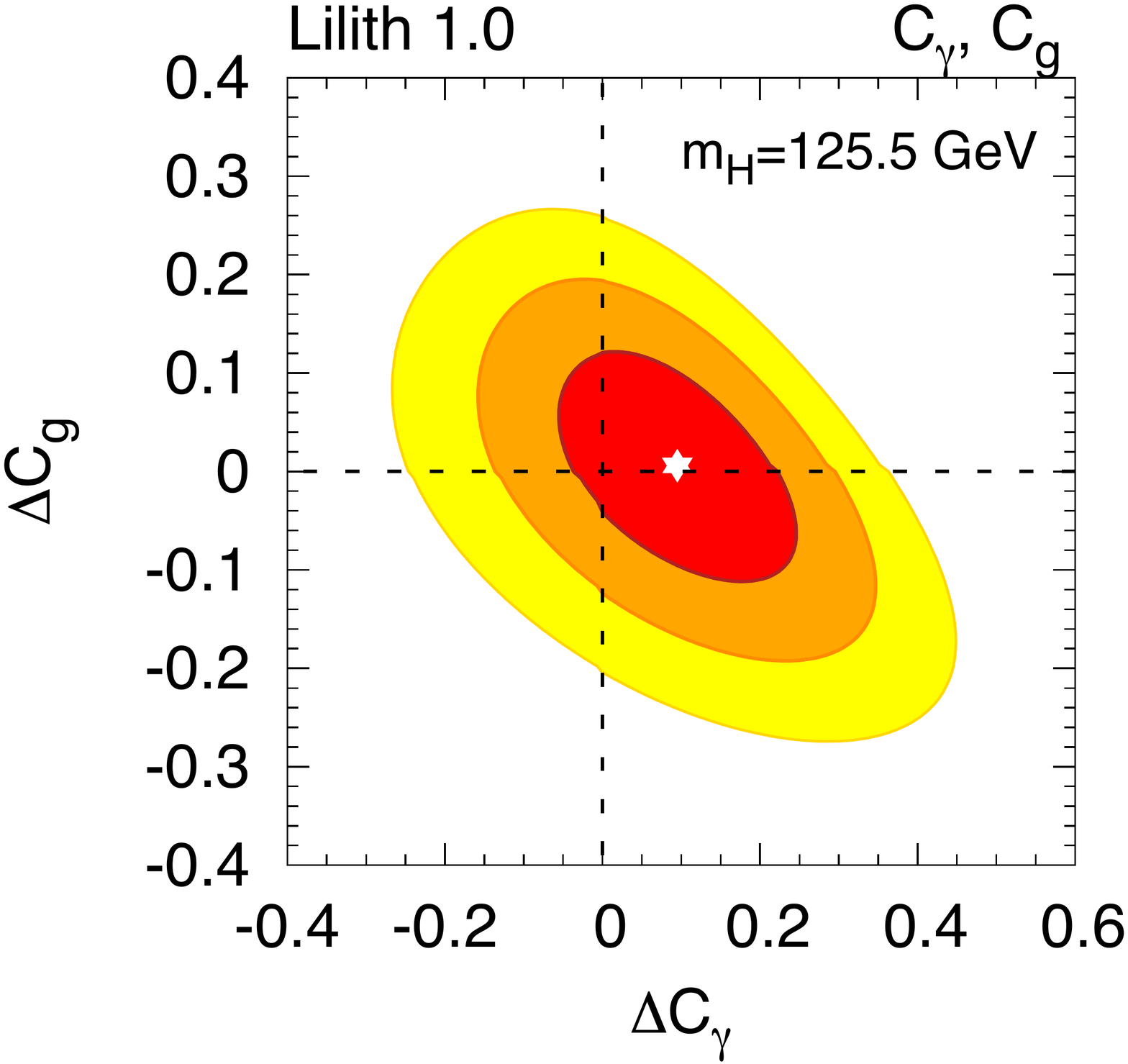}\vspace*{-3mm}
\caption{As Fig.~\ref{fig:cu-cd-cv} but for a 2-parameter fit of $C_g$ and $C_\gamma$ with $\cu=\cd=\cv=1$; 
$\dcg=\cg-1$, $\dcp=\cp-1$. 
\label{fig:cg-cp} }
\end{figure}

So far, we considered deviations of the tree-level reduced couplings from unity, but no extra loop contributions to the effective couplings to gluons and/or photons. If instead we set $C_{U,D,V}=1$ but allow $\cg$ and $\cp$ to vary freely,  
corresponding to loop contributions $\dcg=\cg-1$, $\dcp=\cp-1$ from new physics, we obtain the result shown in Fig.~\ref{fig:cg-cp}. 
The best-fit point has $\dcg=-0.01$ and $\dcp=0.09$, as expected from Fig.~\ref{fig:ellipses1}. 
Again, the SM solution $\dcg=\dcp=0$ lies within the $1\sigma$ contour.

\clearpage 
The current status of invisible (unseen) decays is as follows: (all limits at 95.4\%~CL)
\begin{itemize}
\item for SM-like couplings,  $\brinv<0.12$ ($\brnew<0.09$);
\item for $C_{U,D,V}=1$ but $\cg$, $\cp$ free,  
we find $\brinv<0.24$ ($\brnew<0.23$); 
\item for free $\cu,\ \cd,\ \cv$ but $\cv<1$, we find $\brinv<0.24$ ($\brnew<0.22$);
 this increases to $\brinv<0.34$ when $\cv$ is unconstrained 
 (in this case no limit on $\brnew$ can be obtained~\cite{Belanger:2013xza}). 
\end{itemize}

\section{Two-Higgs-Doublet Models}

In view of the discussion above it is clear that models with an extended Higgs sector will  be significantly constrained by the data. In particular, it is interesting to consider the simplest such extensions of the SM, namely Two-Higgs-Doublet Models (2HDMs) of Type~I and Type~II. 
The basic parameters describing the couplings of the neutral Higgs states to SM particles  
are only two: the CP-even Higgs mixing angle $\alpha$ and the ratio of  the vacuum expectation values,  
$\tanb=v_u/v_d$. 
The couplings, normalized to their SM values, of the Higgs bosons to vector bosons ($C_V$) 
and to up- and down-type fermions ($C_U$ and $C_D$) are functions of $\alpha$ and $\beta$ 
as given in Table~\ref{tab:2hdm-couplings}; see {\it e.g.}~\cite{Gunion:1989we} for details. 
The Type~I and Type~II models are distinguished only by the pattern of their fermionic couplings. 

\begin{table}[h!]
\begin{center}
\begin{tabular}{|c|c|c|c|c|c|}
\hline
\ & Type I and II  & \multicolumn{2}{c|}  {Type I} & \multicolumn{2}{c|}{Type II} \cr
\hline
Higgs & $C_V$ &  $C_U$ &  $C_D$ & $C_U$ &  $C_D$ \cr
\hline
 $h$ & $\sin(\beta-\alpha)$ & $\cosa/ \sinb$ & $\cosa/ \sinb$  &  $\cosa/\sinb$ & $-{\sina/\cosb}$   \cr
\hline
 $H$ & $\cos(\beta-\alpha)$ & $\sina/ \sinb$ &  $\sina/ \sinb$ &  $\sina/ \sinb$ & $\cosa/\cosb$ \cr
\hline
 $A$ & 0 & $\cotb$ & $-\cotb$ & $\cotb$  & $\tanb$ \cr
\hline 
\end{tabular}
\end{center}
\vspace{-.15in}
\caption{Tree-level couplings $\cv$, $\cu$, $\cd$ for the two scalars $h,\ H$ and the pseudoscalar $A$ 
in Type~I and Type~II 2HDMs; $\sina\equiv\sin\alpha$, $\cosa\equiv\cos\alpha$, $\sinb\equiv\sin\beta$, $\cosb\equiv\cos\beta$.}
\label{tab:2hdm-couplings}
\end{table}

To investigate the impact of the current Higgs data on 2HDMs, we vary 
$\alpha = [-\pi/2,\,+\pi/2]$ and $\beta = [0,\,\pi/2[$. 
(Note that this results in $C_U>0$ in our convention, while $C_V$ can be negative).
We implicitly assume that there are no contributions from non-SM particles to the loop 
diagrams for $\cp$ and $\cg$.  In particular, this means our results correspond to the case where the charged 
Higgs boson, whose loop might contribute to $\cp$, is heavy.  

\begin{figure}[t!]\centering
\includegraphics[width=6.5cm]{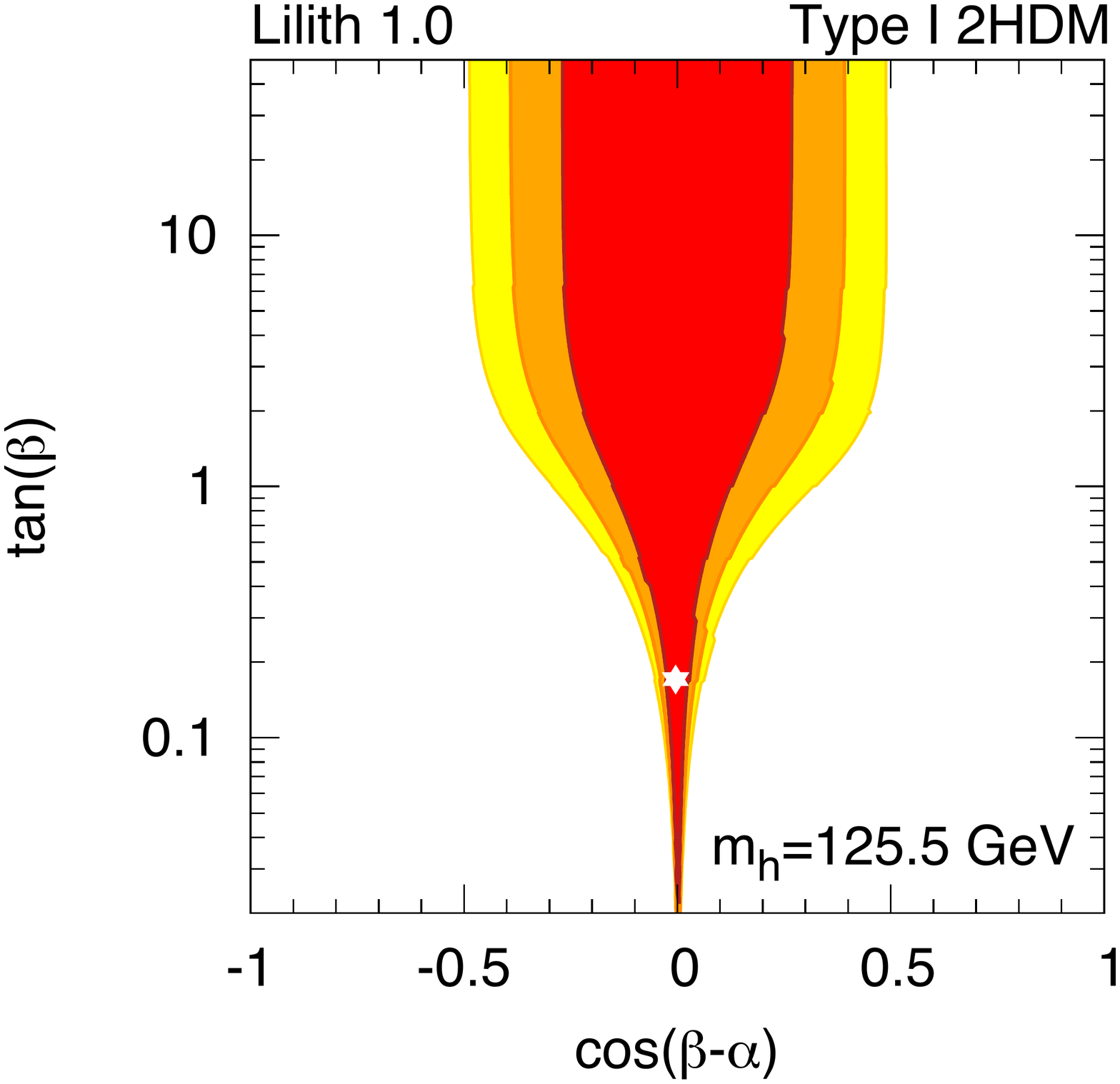}\hspace*{-1cm} 
\includegraphics[width=6.5cm]{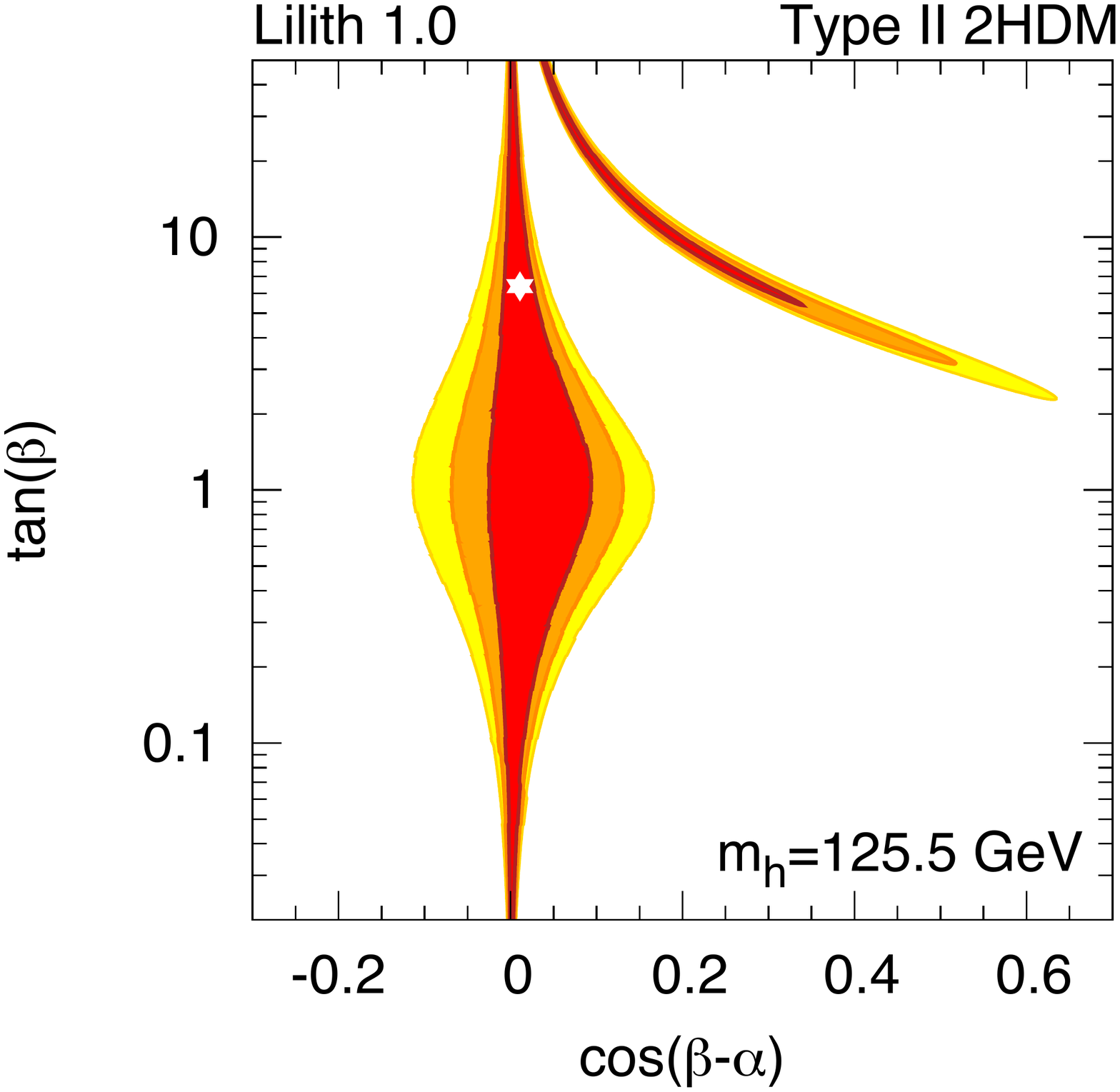}
\caption{Fits of $\cosba$ versus $\tanb$ for the  2HDM of Type~I (left) and of Type~II (right) for $m_h=125.5$~GeV.
The red, orange and yellow areas are the 68.3\%, 95.4\% and 99.7\%~CL regions, respectively.  
The best-fit points are marked as white stars. 
Decays into non-SM particles (such as $h \to AA$) are assumed to be absent.
\label{fig:2hdm} }
\end{figure}

The results of the 2HDM fits are shown in Fig.~\ref{fig:2hdm} for the case that the observed state at $125.5$~GeV
is the lighter CP-even $h$.  
In the case of the Type~I model, we note a broad valley along the SM limit of $\cos(\beta-\alpha)=0$, which is 
rather flat in $\tan\beta$. For $\tan\beta\gtrsim 2$, at 95.4\%~CL $|\!\cos(\beta-\alpha)|$ can be as large as $\approx 0.4$;  
only for $\tan\beta\ll 1$, one is forced into the decoupling/alignment regime. 
The situation is quite different for the Type~II model. Here we observe two narrow valleys in the 
$\tan\beta$ versus $\cos(\beta-\alpha)$ plane. The first one lies along the SM solution $\cosba=0$; 
the largest deviation here occurs around $\tan\beta\approx 1$, where $\cos(\beta-\alpha)\approx 0.13$ 
is allowed at 95.4\%~CL; for both $\tan\beta\gg 1$ and $\tan\beta\ll 1$ one is forced into the decoupling/alignment regime. 
The second minimum is a banana-shaped valley with $\tan\beta\gtrsim 3$ (5) and 
$\cos(\beta-\alpha)\lesssim 0.35$ (0.5) at 68.3\% (95.4\%)~CL.
This corresponds to the degenerate solution with $\CD\approx-1$. 
In 1D, the 68.3\% (95.4\%)~CL limits are 
$|\!\cos(\beta-\alpha)|<0.19$ $(0.34)$ for Type~I and 
$\cosba=[0, 0.29]$ ($[-0.05,0.47]$) for Type~II; 
the latter shrinks to $\cosba=[0, 0.07]$ ($[-0.05,0.11]$) when demanding $\cd>0$.

Constraints on and future prospects for 2HDMs in light of the LHC Higgs signal (status Spring 2013) 
were discussed in detail in \cite{Dumont:2014wha} taking into account all relevant theoretical and 
experimental constraints. 
The results of that paper will be somewhat modified by the new constraints presented here; 
this is presently under study.

\section{Conclusions}

We presented a brief update of the global fits of the 125.5 GeV Higgs boson using all publicly available experimental results as per Summer 2014. The fits were done with {\tt Lilith~1.0}, a new user-friendly public tool for evaluating 
the likelihood of an SM-like Higgs boson in view of the experimental data. 
Our results can be summarized as follows:
\begin{enumerate}
\item The latest ATLAS and CMS results for the $H\to \gamma\gamma$ decay mode 
now point to a very good agreement with the SM; concretely we get   
$\widehat{\mu}^{\rm{ggF+ttH}}_{\gamma\gamma}=1.25\pm 0.24$ and $\widehat{\mu}^{\rm VBF+VH}_{\gamma\gamma}=1.09\pm0.46$ 
with a correlation of $\rho=-0.30$. 
\item In the $\cu,\cd,\cv$ reduced coupling fit, we found $\cu=1.01\pm0.1$, $\cd=1.01\pm0.16$ 
and $\cv=1.05\pm 0.08$, which leads to $C_g=1.01\pm0.11$ and $C_\gamma=1.06\pm0.11$ (in 1D). 
\item Custodial symmetry can also be tested. We found $\cwz=0.92\pm0.1$, hence compatibility with custodial symmetry at the $1\sigma$ level. 
\item Assuming SM-like couplings, the limit for invisible decays is $\brinv<0.12$ at 95.4\%~CL. 
This changes to $\brinv<0.34$ when $\cu,\cd,\cv$ (or even $\cu,\cd,\cv,\cg,\cp$) are allowed to vary. 
\item In the context of 2HDMs, barring loop contributions from the charged Higgs, the 95.4\%~CL limits in 1D 
are $\sinba>0.94$ in Type~I and  $\sinba>0.90$ in Type~II. 
\end{enumerate} 

As mentioned, one of the limitations of these fits is the use of the Gaussian approximation. This could easily be avoided 
if the experimental collaborations published the 2D likelihood grids in addition to the 68\% and 95\%~CL contours. 
Another limitation is induced by the combination of production modes, typically ${\rm ggF+ttH}$ and ${\rm VBF+VH}$, in the experimental results. This could be overcome if the collaborations provided the signal strength likelihoods beyond 2D projections---the optimum would be to have the signal strengths as functions of $m_H$ separated into all five production modes ggF, ttH, VBF, ZH and WH, as recommended in \cite{Boudjema:2013qla}. 
We hope that this way of presentation (in digital form) will be adopted for Higgs results at Run-2 of the LHC. 
The structure of {\tt Lilith} is well suited to make use of such extended experimental results.

\section*{Acknowledgements}

We thank John F.\ Gunion and Yun Jiang for discussions. 
J.B.\ is supported by the ``Investissements d'avenir, Labex ENIGMASS''.

\bibliography{hfit}  

\end{document}